\documentclass[conference]{IEEEtran}

\usepackage{subfig}
\usepackage{color}
\usepackage[colorinlistoftodos]{todonotes}

\usepackage{url}
\usepackage{xcolor, soul}
\usepackage{tikz}
\usepackage{authblk}
\usepackage{graphicx} 

\title{Evaluating Vulnerability of Chiplet-Based Systems to Contactless Probing Techniques}
\author[1]{Aleksa Deric}
\author[2]{Kyle Mitard}
\author[2]{Shahin Tajik}
\author[1]{Daniel Holcomb}
\affil[1]{Department of Electrical and Computer Engineering, University of Massachusetts, Amherst}
\affil[ ]{\textit {\{aderic, dholcomb\}@umass.edu}}
\affil[2]{Department of Electrical and Computer Engineering, Worcester Polytechnic Institute}
\affil[ ]{\textit {\{krmitard, stajik\}@wpi.edu}}

\date{April 2024}

\usetikzlibrary{positioning}
\usetikzlibrary{calc} 

\newcommand{\degrees}[1] {{$#1^{\circ}{\rm C}$}}

\sethlcolor{yellow}

\begin{document}

\maketitle

\begin{abstract}
Driven by a need for ever increasing chip performance and inclusion of innovative features, a growing number of semiconductor companies are opting for all-inclusive System-on-Chip (SoC) architectures. Although Moore's Law has been able to keep up with the demand for more complex logic, manufacturing large dies still poses a challenge. Increasingly the solution adopted to minimize the impact of silicon defects on manufacturing yield has been to split a design into multiple smaller dies called chiplets which are then brought together on a silicon interposer. Advanced 2.5D and 3D packaging techniques that enable this kind of integration also promise increased power efficiency and opportunities for heterogeneous integration.

However, despite their advantages, chiplets are not without issues. Apart from manufacturing challenges that come with new packaging techniques, disaggregating a design into multiple logically and physically separate dies introduces new threats, including the possibility of tampering with and probing exposed data lines. In this paper we evaluate the exposure of chiplets to probing by applying laser contactless probing techniques to a chiplet-based AMD/Xilinx VU9P FPGA. First, we identify and map interposer wire drivers and show that probing them is easier compared to probing internal nodes. Lastly, we demonstrate that delay-based sensors, which can be used to protect against physical probes, are insufficient to protect against laser probing as the delay change due to laser probing is only 0.792ps even at 100\% laser power.

\end{abstract}

\section{Introduction}
\label{sec:intro}
Chiplets are separately produced silicon dies that are assembled on an interposer to form a composite system that is comparable to a monolithic integrated circuit. With chip designers squeezing in more logic and functionality each generation, chiplets are becoming more common. Chiplets allow designers to circumvent reticle limits and provide an opportunity to increase yield, leverage multiple process nodes for heterogeneous integration, and reuse IP to cut down on costs. Driven by demand for CPUs and GPUs in data centers, by 2027 the market for chiplet-based processors is expected to reach \$135 billion~\cite{chipletsummit2023}. Currently all the major semiconductor players including AMD~\cite{amd2022rdna3}, Intel~\cite{intel2023meteorlake}, Nvidia~\cite{nvidia2022nvlink} and Apple~\cite{apple2022m1ultra} feature chiplet-based products in their lineup. Due to the extra effort needed to design and package chiplets, they have initially proliferated in top of the line products that can absorb the added overhead costs. More recently however, there's been an effort to democratize the technology by allowing users to configure their own SoCs to be assembled from a library of pre-made chiplets built using open standards~\cite{zeroasic2024composable}.

Unfortunately, disaggregating a system into parts opens a lot of new attack vectors, including possibility of cross-die side-channels, introduction of hardware Trojans in the supply chain, IP piracy, swapping of individual chiplet dies, and probing of chiplet interfaces. Among these attack vectors, probing at chiplet interfaces is of particular concern, as chiplets are split by functionality during floorplanning for timing closure reasons, and therefore signals carried on the interposer wires may be important data channels that are of high-value to an attacker.

To the best of our knowledge, this paper is the first work to demonstrate contactless probing techniques in the context of chiplets, wherein the large drivers for long wires across the passive interposer are a uniquely exposed target. The specific contributions of this paper are:

\begin{itemize}
\item We experimentally demonstrate, for the first time, a laser probing attack against chiplets, which are fabricated in 16nm technology and packaged with a silicon interposer in 65nm technology.
\item We compare the exposure of inter-chiplet wires relative to that of on-die wires in the same technology.
\item We assess the effectiveness of delay-based sensors in detecting contactless probing attacks against chiplets.
\end{itemize}

The remainder of the paper is organized as follows: Section~\ref{sec:background} provides background on chiplets and failure analysis. Sections~\ref{sec:laser-probing-techniques} and~\ref{sec:experimental-setup} explain the techniques and experimental setup used in the paper. Section~\ref{sec:laser-probing} demonstrates probing of die-to-die connections and compares their vulnerability to that of on-die connections. Finally, in Section~\ref{sec:delay-sensors} we attempt to detect probing using two types of delay sensors and discuss the limitations of effective sensor-based mitigation against laser probing.

\section{Background}
\label{sec:background}
In this section, we discuss the defining features of chiplet interconnects and cover relevant previous work in the areas of chiplet security and failure analysis. Our work covers a unique gap at the intersection of these two areas, pertaining to probing of chiplet interfaces. 

\subsection{Chiplet Interconnect Technology}
\label{subsec:chiplets}
The distinguishing feature of chiplets is their use of die-to-die interconnect technologies that make chiplet-based systems significantly more performant than systems comprising multiple chips integrated on a PCB. Signals between chiplets travel a distance of only a few mm, through microbumped connections and densely packed wires of a multi-layer interposer, as depicted in Figure~\ref{subfig:3d-chiplet-overview}. Although the wires in die-to-die connections are much smaller than wires used in board-level connections, they have significant capacitance relative to local on-die wires routed in lower metal layers, and therefore are driven with upsized transistors; probing these upsized transistors is the focus of this paper.

Intel~\cite{kehlet2017accelerating}, TSMC~\cite{lin2019}, and AMD/Xilinx~\cite{saban2012xilinx} all use their own chiplet interfaces, and UCIe~\cite{ucie-spec} is emerging as a standard interface to enable interoperability of chiplet parts from different vendors. Though different in some implementation details, all the listed interfaces provide capability for sub-pJ energy-per-bit and a 35$\mu$m to 55$\mu$m bump pitch with hundreds to thousands of wires per mm of shoreline to deliver hundreds of GB/s of aggregate bandwidth. Other shared features include low-resistance microbump connections to the interposer, shielded wiring and source-synchronous clocking. In a source-synchronous system the transmitting clock is forwarded along with the data lines, and the receiving chiplet de-skews the forwarded clock to reliably sample the incoming data. All of these features make chiplets highly attractive when building high-performance SoCs.

\subsection{Related Work in Chiplet Security}
\label{subsec:chiplet-security}
With fast progress being made toward addressing reliability, performance and testing~\cite{hutner2020special}\cite{abdennadher2021testing} challenges, security of chiplets has been left as an afterthought. Disaggregation of a system into multiple smaller dies opens a possibility for new side-channel attacks, hardware Trojans or counterfeits in the supply chain, probing of chiplet interfaces, reverse engineering, and even swapping of individual chiplet dies. 
Several papers have surveyed and raised awareness of these possible issues~\cite{vashistha2022toshi}\cite{khansecurity}.

There are few concrete examples of real word attacks on chiplets, though in recent years research demonstrating attacks has started to appear in open literature. \cite{khan2023exploring} proposes a reverse engineering framework for chiplet packages and demonstrates its effectiveness by using 3D X-ray tomography to extract information about the package architecture which could be used to recover the netlist and pin mappings. Similarly, \cite{wang2019reverse} discusses a SAT-based attack to reverse engineer missing connections of 2.5D split manufacturing netlists. Finally,~\cite{cross-slr-covert}\cite{mustafa2023covert} show that an inter-chiplet covert channel can be created through a shared power distribution network (PDN) that is present in many chiplet architectures.

Likewise, various solutions have been proposed to defend against attacks including using an active interposer~\cite{chacon2021coherence}\cite{nabeel2020} to police traffic between chiplets and a dedicated security chiplet that actively monitors power traces~\cite{zhang2023sipguard}. Lastly, treating individual chiplets as their own secure systems in an otherwise insecure environment means the existing cryptographic techniques can be applied as an additional layer of protection on top of the interconnect~\cite{ceva-product-note}. Still, many of the attack vectors frequently mentioned remain theoretical and more work is needed to validate the threat models discussed in literature.

Importantly, the existing works on chiplet security have not yet considered the problem of laser probing attacks. The vast number of die-to-die connections, and the physical characteristics of the die-to-die circuits, render them considerably exposed to an adversary that performs laser probing. This attack is the focus in our paper.

\subsection{Failure Analysis Applied to Security}

In order to refine process parameters and increase yield, various failure analysis methods have been developed. While larger geometries can be analyzed with visual inspection or decapsulation and microprobing, smaller geometries require more delicate and precise techniques. Advanced imaging, optical microscopy, and e-beam probing techniques have been developed for this purpose. Focused Ion Beam (FIB) is one example of an invasive technique for analyzing and isolating faults. Less destructive techniques are desirable when preserving the device-under-test (DUT) is of importance, or the target data being extracted requires the DUT to be running and functional. Near-infrared (NIR) wavelength photons are able to pass through silicon and can thus be used to non-destructively peek through the back side of dies. This is made especially relevant with flip-chip packages being a common way of maximizing performance and I/O pin count. And although this backside technique is limited to probing surface structures such as transistors and is unable to look deeper into the die or past the metal layers, reading out values from individual transistors is sufficient for using the technique offensively.

The signal-to-noise ratios (SNR) of optical and e-beam probing are compared in recent work~\cite{optical-vs-ebeam}. Research has also considered how to deploy optical probing to search for, and find, signals of interest in a large circuit~\cite{transistor-jungle}\cite{sass2023modulation}. More critically,~\cite{tajik-ccs17} demonstrated how optical contactless techniques operating with light sources in the NIR region can be used to bypass on-die encryption and extract a full plaintext bitstream from a 28nm AMD/Xilinx FPGA. Similarly, \cite{lohrke2018key} showed how thermal laser stimulation (TLS) can be used to read out secret keys from inside a battery-backed RAM (BBRAM) of an FPGA - the process which can even be fully automated~\cite{transistor-jungle} to further reduce the cost and time required to conduct such an attack.

Until now, there are few works that have employed probing against a 16nm technology, and no works that have evaluated probing against inter-chiplet connections. Our work in this paper fills this gap in knowledge by experimentally validating the threat of contactless probing in the context of chiplets and demonstrating that delay-based sensors cannot provide adequate protection.

\subsection{Threat Model}
Our threat model assumes that the attacker can interact with and has physical possession of a multi-die flip-chip device from which they wish to extract secret data such as decryption keys. The attacker rents a contactless probing system from a failure analysis lab with the rest of their arsenal consisting of off-the-shelf tools and commodity hardware. As discussed in the introduction, a chiplet system is likely to be partitioned based on functionality of individual modules - this is especially true if individual chiplets are designed by a third party and merely used as IPs - therefore, the quickest path to success might involve locating and probing communication links going between the chiplets. Although we use an FPGA as our target in our work which lets us flash different bitstreams, in a real-world scenario the adversary could initially perform profiling steps on a training device before moving to perform a probing attack on the actual victim device. Similarly, given an ASIC design that is not reconfigurable, the profiling process might involve comparing and locating identical structures using optical imaging and manipulating input data to induce and detect specific frequencies with techniques like Electro-Optical Frequency Mapping (EOFM). Despite no custom reconfiguration ability, the steps to profile an ASIC are analogous to the steps we take in the paper.

\section{Contactless Probing Techniques}
\label{sec:laser-probing-techniques}
In this section, we provide as background an overview of the contactless probing techniques employed in the paper, and explain the working principle behind each one. A detailed description of how the techniques are specifically applied in our experiments is found in Section~\ref{sec:laser-probing}.

\subsection{Photon Emission}
\label{subsec:photon-emission}
Measuring photon emission of a chip provides coarse information about the activity level of circuits at different locations on the chip. Complementary metal-oxide semiconductor (CMOS) gates that are the backbone of all modern digital integrated circuits (ICs) are constructed by putting together complementary pairs of nMOS and pMOS transistors. When a CMOS gate is quiescent, that is none of its nodes are switching, there are only small leakage currents flowing through its transistors, and hence almost no photons are being emitted. However, when an input to the gate changes, and the change results in the output node changing too, large currents flow to charge or discharge individual node capacitances as well as in the form of short-circuit current. The energy of accelerated hot carriers traveling through transistor channels is sometimes released as photons, which travel though the silicon backside of the IC and can be captured by a photodetector and integrated across time to construct an activity map of the device.

\subsection{Electro-Optical Frequency Mapping (EOFM)}
\label{subsec:eofm}
Electro-Optical Frequency Mapping (EOFM) or laser voltage imaging (LVI) is a laser probing technique in which a set of precisely controlled mirrors scan a laser beam across the device and build a detailed activity map of nodes switching. More specifically, after leaving the light source, the laser beam travels through the silicon and is reflected off the metal structures on the silicon surface. While it is scanned, the laser is held at each position for a period of time, called dwell time, and the returning beam is fed into a spectrum analyzer with a filter that passes a specific frequency band of interest. Nodes in a circuit that switch at a particular frequency will modulate the reflected light at their switching frequency. The bandpass filter of the spectrum analyzer therefore allows for detecting activity of nodes that are switching at specific frequencies of interest, and rejecting activity at other frequencies. The measurements collected during the entire scan can be used to build a frequency-selective activity map.

\subsection{Electro-Optical Probing (EOP)}
\label{subsec:eop}
\begin{figure}
\centering
\includegraphics[width=0.48\textwidth]{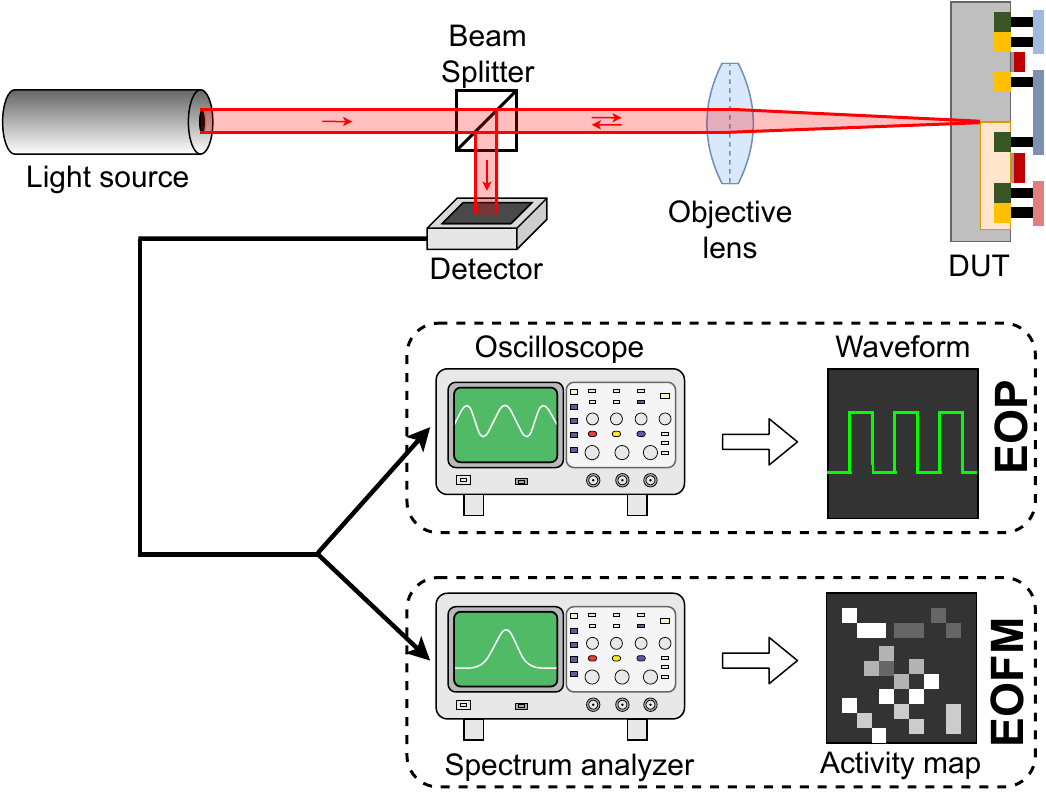}
    \caption{Diagram illustrating the basic working principles of EOP and EOFM, as described in Sections~\ref{subsec:eop} and ~\ref{subsec:eofm} respectively. Apart from EOFM involving scanning the beam, the two techniques differ in how the reflected light is processed. }
    \label{fig:eop-eofm-principle}
\end{figure}

Electro-Optical Probing (EOP) or laser voltage probing can be thought of as a variation of more general Laser Voltage Probing (LVP) techniques where an incoherent light source in the former is substituted for the laser typically used in the latter. EOP works by focusing light on a single point of interest one wishes to probe and measuring the intensity of the reflected beam over time. The degree to which the returning light's intensity is altered depends on carrier concentrations in the area of interest i.e., electrical signal at the targeted spot. Hence by feeding the returning beam into a photodetector and measuring its intensity the voltage level of the node being probed can be determined. The intensity difference is however small enough that multiple measurement repetitions must be collected and averaged to obtain a signal with an acceptable signal-to-noise ratio. To align the repeated measurements, a trigger reference synchronized to the target waveform must be provided. Figure~\ref{fig:eop-eofm-principle} illustrates EOFM and EOP. Note that the two techniques are similar apart from how the returning beam is processed.

\section{Experimental Setup}
\label{sec:experimental-setup}
In this section, we introduce our experimental setup, including the chiplet-based FPGA that acts as our probing target and the failure analysis microscope used to execute the attacks. We also briefly describe the board preparation process and the data collection framework.

\subsection{Chiplet FPGA Platform}
\label{subsec:vcu118}
AMD/Xilinx has offered chiplet-based FPGAs since 2011 under the name of Stacked Silicon Interconnect~(SSI) devices~\cite{saban2012xilinx}. In AMD/Xilinx architecture, the chiplets are called Super Logic Regions (SLRs) and the die-to-die connections, which they call Super Long Lines (SLLs), are routed through low-resistance microbumps and a passive silicon interposer. AMD/Xilinx chiplets do not implement any particular interface standard; instead, the SLLs simply begin and terminate with special Laguna registers which are clocked using the FPGA's programmable clocking network that is shared across SLRs through dedicated clock spines spanning the entire device. AMD/Xilinx hierarchy organizes groupings of six SLLs into so-called Laguna sites, and groupings of four Laguna sites (24 SLLs) into so-called Laguna tiles.

In this paper, we conduct experiments using the VCU118 development board which features a VU9P FPGA (part number xcvu9p-flga2104-2L-e). The VU9P is a high-capacity UltraScale+ FPGA consisting of three identical 16nm chiplets sitting on a 65nm interposer in a flip-chip configuration with 17,280 SLLs (720 Laguna tiles) connecting each SLR to its neighbors.

\subsection{Board Preparation}
\begin{figure}
    \centering
    \subfloat[]{
        \includegraphics[width=0.21\textwidth]
        {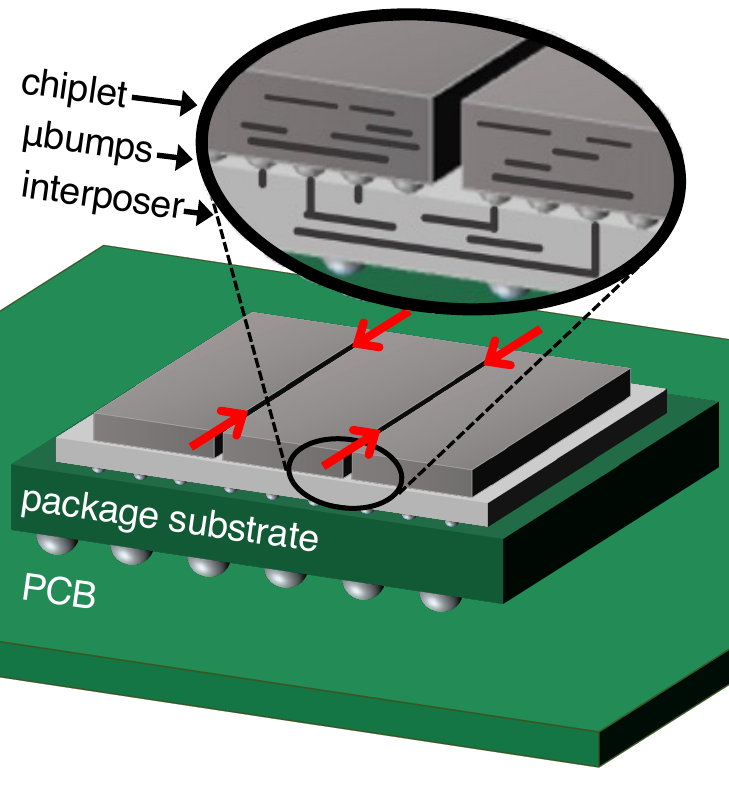}
        \label{subfig:3d-chiplet-overview}
    }
    \hfill
    \subfloat[]{
        \includegraphics[width=0.25\textwidth]
        {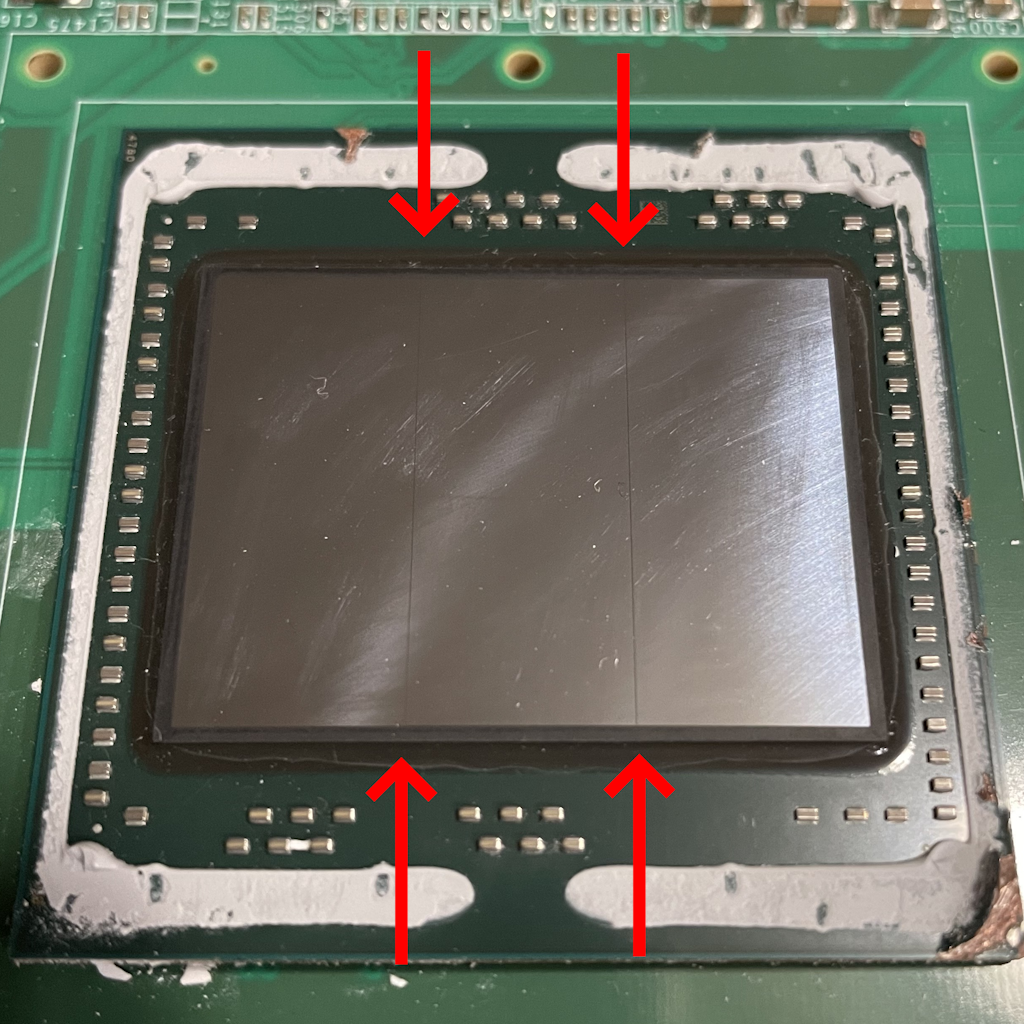}
        \label{subfig:3d-chiplet-photo}
    }    
    \caption{(a)~Diagram of chiplet-to-chiplet connections through interposer, and (b)~corresponding photograph of delidded AMD/Xilinx chiplet-based VU9P FPGA used in evaluation. The red arrows in each picture annotate the chiplet boundaries.}
    \label{fig:overview-diagram}
\end{figure}

Optical probing requires direct access to the backside silicon of the FPGA, so the board must be prepared to expose this surface. Since the VCU118 board has a large copper heatsink and fan to dissipate the heat generated by the FPGA, the first step is to remove the heatsink. Next, the Integrated Heat Spreader (IHS) is removed by inserting a chisel between it and the package substrate that connects to the PCB, and carefully applying force to break the two loose. Finally, we remove a thin white coat of thermal paste covering the chiplet dies, by wiping it away with a 99.9\% isopropyl alcohol solution. The chiplet backsides revealed by this preparation are shown in Figure~\ref{subfig:3d-chiplet-photo}.

\subsection{Command and Control}
\begin{figure}
    \centering
    \includegraphics[width=0.48\textwidth]{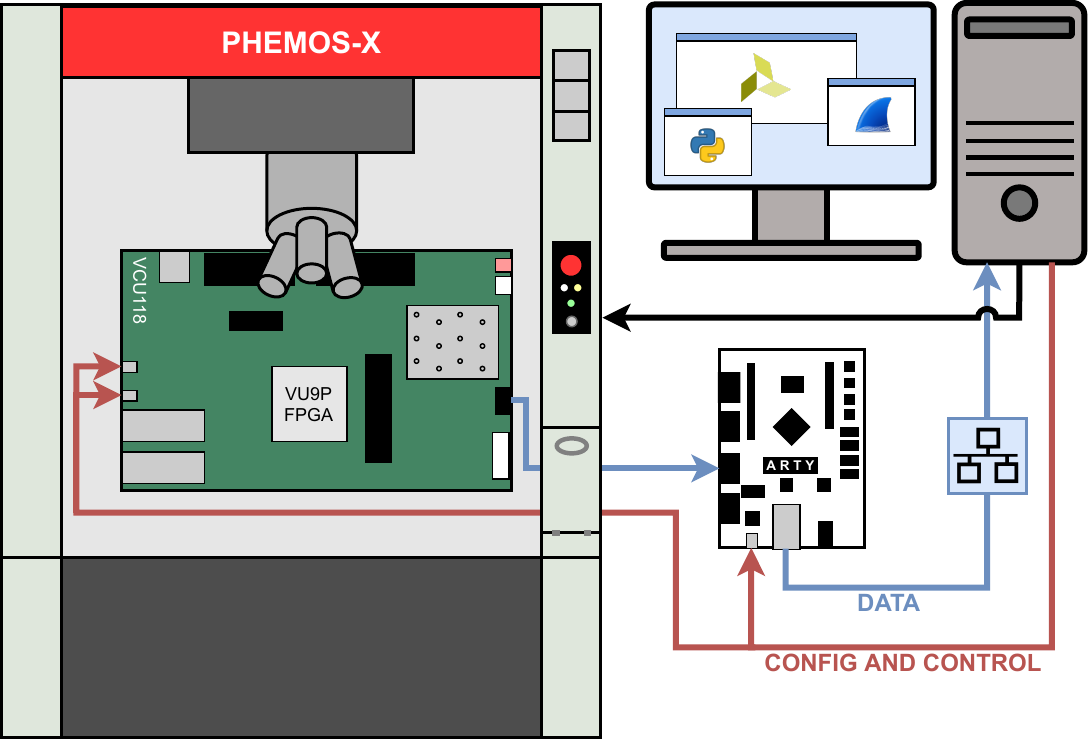}
    \caption{Illustration of the setup used for data collection.}
    \label{fig:exp-setup}
\end{figure}

Without any mechanism to help the FPGA dissipate heat, we observe that idle temperatures hover between \degrees{65} and \degrees{75}, almost \degrees{40} higher than with the heat sink. To avoid overheating the FPGA we use a second board, an Arty A7-T35 (part number xc7a35ticsg324-1L), to offload much of the data collection logic (see Figure~\ref{fig:exp-setup}). Samples generated on the VU9P are immediately placed in a shift register that serially forwards them over PMOD~\cite{digilent2017} ports to the Arty where they are buffered and packetized before being sent over Ethernet. Offloading the processing logic also has the added benefit of reducing on-chip switching noise during measurement.

\subsection{Emission and Laser Scanning Microscope}
Probing experiments are performed using the Hamamatsu PHEMOS-X Emission and Laser Scanning Microscope. The microscope features an InGaAs camera for photon emission and a 1.3$\mu$m wavelength laser for optical probing~\cite{hamamatsu2022}. A set of galvanometric mirrors guides the laser while the reflected light can be collected using one of the four objective lenses available: 5x/0.14~NA, 20x/0.4~NA, 50x/0.76~NA and 71x/0.86~NA~\cite{hamamatsu2022}. The microscope is controlled by a desktop computer running the Hamamatsu control software.

After preparing the FPGA, the board is placed under the microscope apparatus inside the PHEMOS-X chamber and the power and communication cables running to the outside are clamped to the frame to prevent movement during experiments. The full experimental setup is pictured in Figure~\ref{fig:exp-setup}.

\section{Laser Probing of Interconnect Drivers}
\label{sec:laser-probing}
The large size of the transistors that drive die-to-die connections makes them an easy and attractive target for probing. These transistors are upsized to supply the current necessary to drive the load of the microbump and the long interposer wire, which may be several mm in length. In the following section we locate a group of drivers and compare, for reference, their probing exposure to the equivalent logic function implemented using on-chip fabric registers (see Figure~\ref{fig:slls-vs-wires}). Note that the off-chip driver is significantly more visible on account of its upsized transistors.

\subsection{Localizing Circuits-of-Interest}
\label{subsec:results-EOFM}

\subsubsection{Using photon emission to navigate the chip}
\label{subsubsec:photon-emission-slls-vs-wiresw}
\begin{figure}[h]
    \centering
    \subfloat[5x]{
        \includegraphics[width=0.47\linewidth]{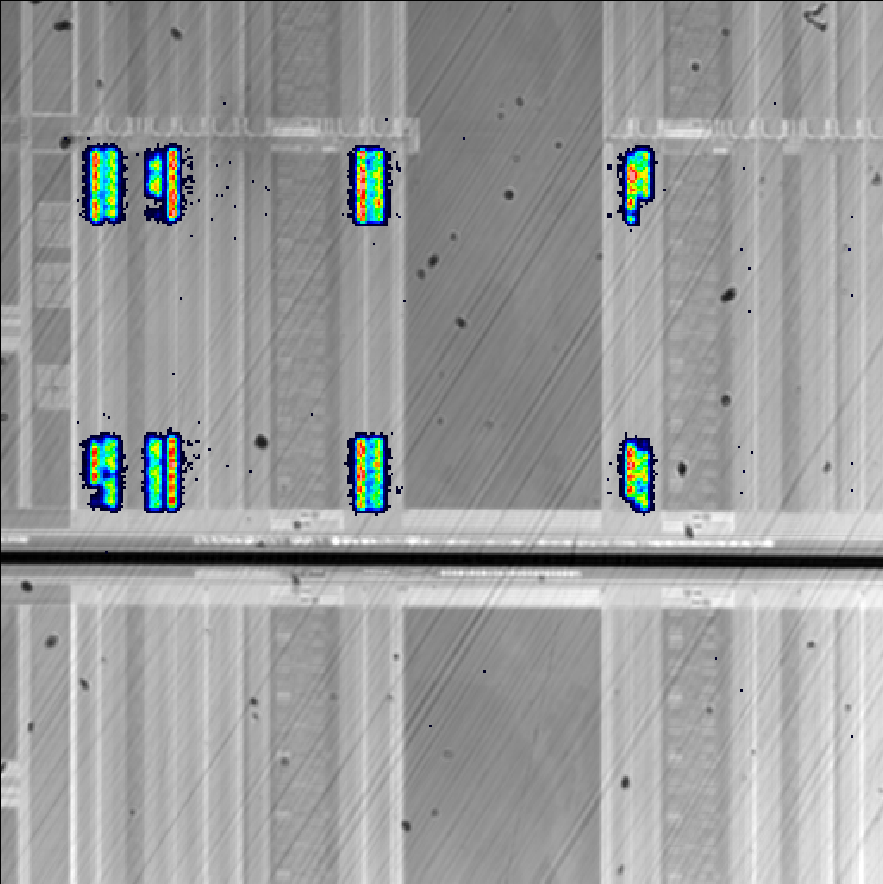}
        \label{subfig:photon-emission-overlaid}
    }
    \subfloat[71x]{
        \includegraphics[width=0.47\linewidth]{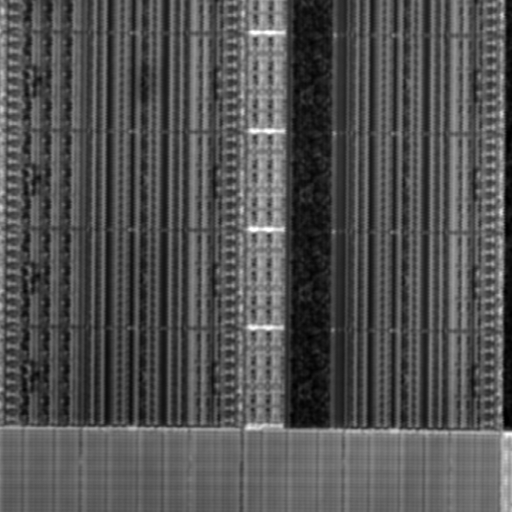}
        \label{subfig:laguna-columns}
    }
    
    \begin{tikzpicture}[remember picture, overlay]
        \draw[blue,thick] (-3.35,2.1) rectangle (-3.0,2.6);
        \draw[blue,ultra thick] (0.15,0.45) rectangle (4.3,4.6);

        \draw[blue,thick,-] (-3.35,2.6) -- (0.15,4.6) {};
        \draw[blue,thick,-] (-3.35,2.1) -- (0.15,0.45) {};

        \draw[green,thick] (2.1,1.95) rectangle (2.45,2.75);
        \draw[red,ultra thick,->] (0,0) ++(-4.44,1.85) -- ++(0.7,0) node[left] {};
        \draw[red,ultra thick,<-] (0,0) ++(-0.66,2.1) -- ++(0.7,0) node[left] {};
    \end{tikzpicture}
    
    \caption{Photon emission highlights circuits with high switching activity, such as groups of custom placed ring oscillators in (a), which are used as guideposts to locate circuits of interests like a Laguna column in (b). A single Laguna tile is outlined in green. The red arrows annotate the boundary between two chiplets.}
    \label{fig:photon-emission-and-columns}
\end{figure}
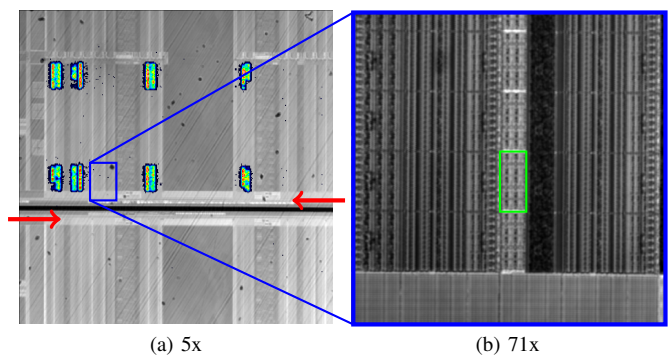

The assembled silicon package contains billions of transistors and measures approximately 35mm$\times$26mm, with each chiplet being 11.66mm$\times$26mm. Navigating the large chip area to find features of interest in the layout can prove challenging. For this purpose, we flash the FPGA with a dummy bitstream with individually controllable blocks of oscillators (256 ring oscillators each) custom placed around the chip to serve as guideposts for learning the physical position of on-die circuits-of-interest. As photon emission can be used under low magnification, we use the 5x lens with 10 second integration time for navigation. Figure~\ref{subfig:photon-emission-overlaid} shows a photon emission view of active ring oscillators overlaid on top of a near-infrared image of the back side of the die. Given that the layout presented in Vivado closely matches the actual layout of the silicon, turning specific groups of oscillators on and off while monitoring photon emission lets us steer the camera to the edge of the chip and to the vicinity of a specific Laguna tile of interest.

\subsubsection{Using EOFM to pinpoint individual cells}
\label{subsubsec:locating-drivers-eofm}

\begin{figure*}[htbp]
    \centering
    \begin{tikzpicture}
        \node[anchor=south west,inner sep=0] (image) at (0,0) {\includegraphics[width=0.99\linewidth]{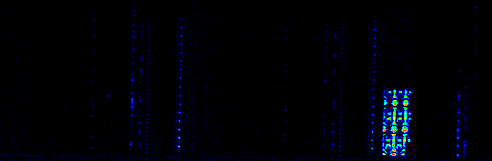}};

        \draw[green,ultra thick] (13.95,0.124) rectangle (15.05,2.63);
        
        \draw[yellow,ultra thick,<-] (image.south west) ++(6.6,0.5) -- ++(1,0) node[right] {};
        \node[align=center, text=yellow] at (8.8,0.5) {Fabric registers};
        
        \draw[green,ultra thick,->] (image.south west) ++(13.2,1.5) -- ++(1,0) node[left] {};
        \node[align=center, text=green] at (12.0,1.5) {Laguna drivers};

    \end{tikzpicture}
    \caption{EOFM of 32 fabric flip-flops and 24 Laguna registers toggling at 100MHz under 20x magnification. The spot pointed to by the yellow arrow represents a group of two registers.}
    \label{fig:slls-vs-wires}
\end{figure*}
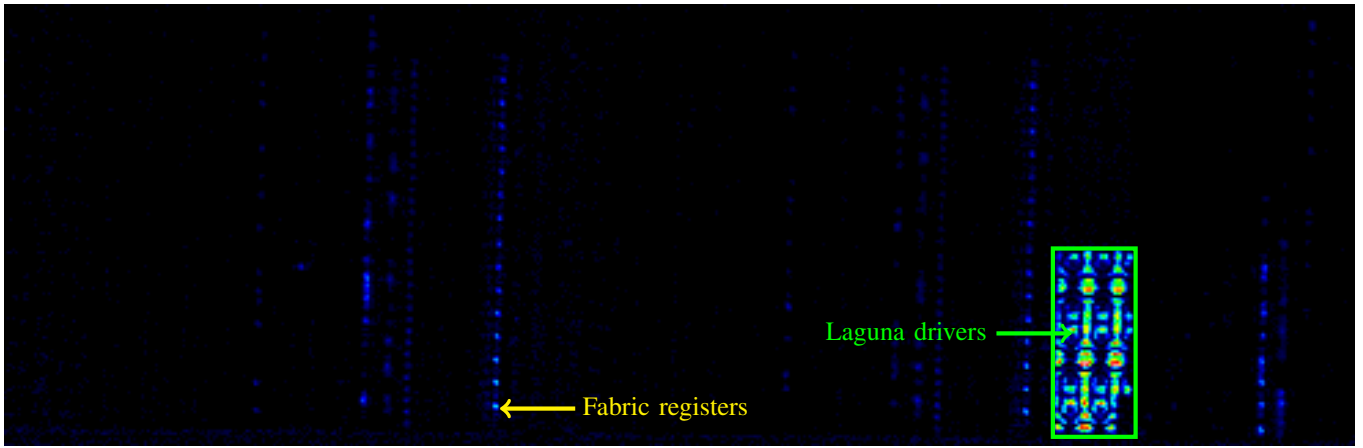

Once in the general area of interest we use EOFM with a higher magnification lens to pinpoint the location of individual registers and drivers. Recall (see Section~\ref{subsec:eofm}) that EOFM shows activity of nodes of interest that toggle at a specific frequency, which in our case is 100MHz. To compare the visibility of fabric registers to Laguna registers, we create a test circuit that drives a 100MHz square wave into 32 fabric registers (corresponding to 2 logic slices) and 24 Laguna registers (corresponding to a single Laguna tile). Figure~\ref{fig:slls-vs-wires} shows the resulting activity map of this basic EOFM scenario under 20x magnification. Although fewer in number, the SLL drivers appear much larger than the slice registers, making them uniquely exposed to probing. Notable is also that each spot on the left side of Figure~\ref{fig:slls-vs-wires} represents not only one, but two slice registers. 

\begin{figure}[h]
    \centering
    \subfloat[]{
        \centering
        \includegraphics[width=0.22\textwidth]{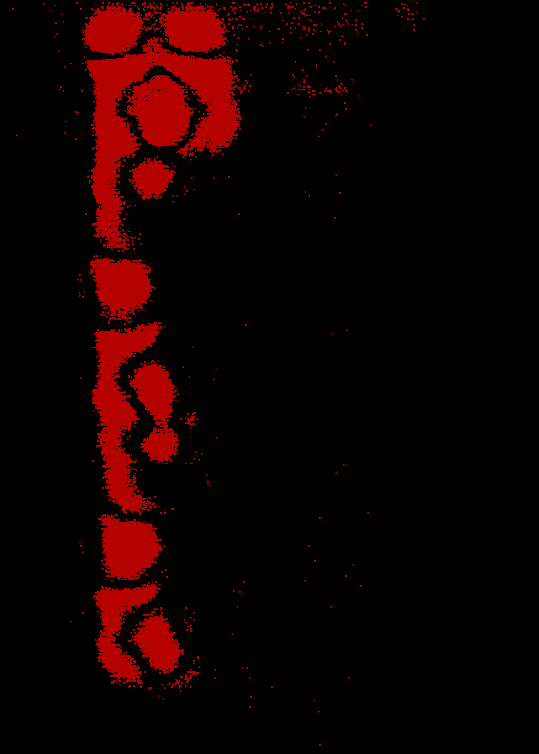}
        \label{subfig:single-laguna-site-layout}
    }
    \hfill
    \subfloat[]{
        \centering
        \includegraphics[width=0.22\textwidth]{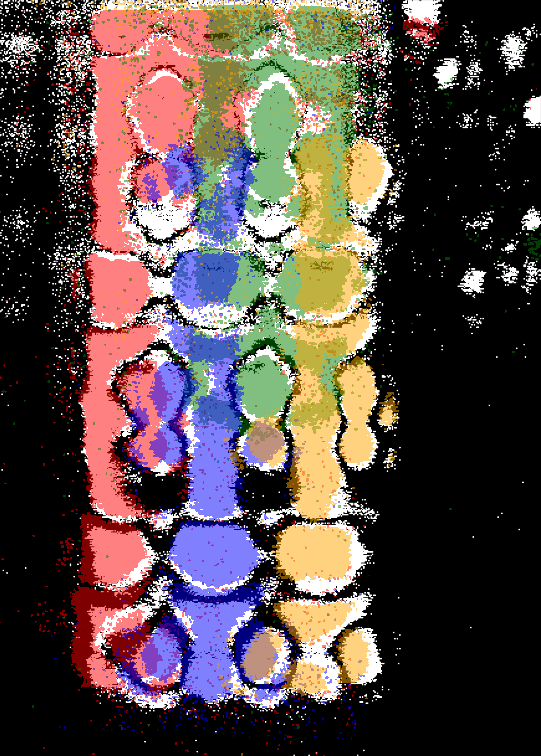}
        \label{subfig:laguna-site-layout}
    }
    \vspace{0.0cm}
    
    \subfloat[]{
        \centering
        \includegraphics[width=0.22\textwidth]{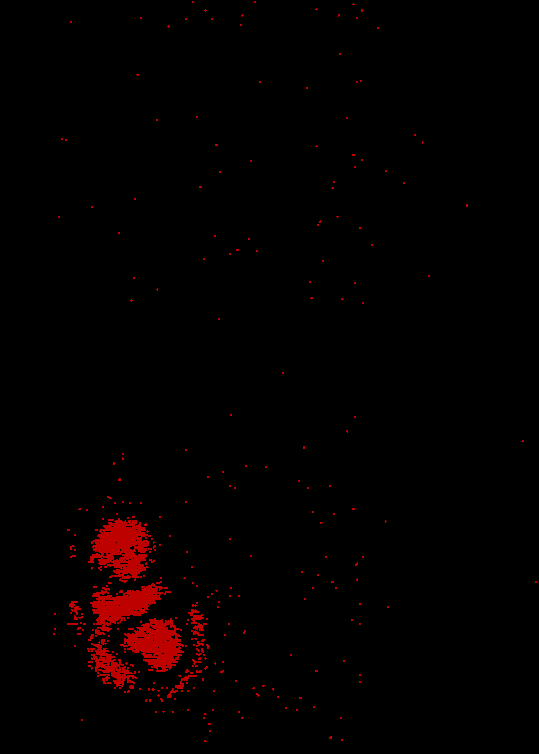}
        \label{subfig:single-laguna-sll-layout}
    }
    \hfill
    \subfloat[]{
        \centering
        \includegraphics[width=0.22\textwidth]{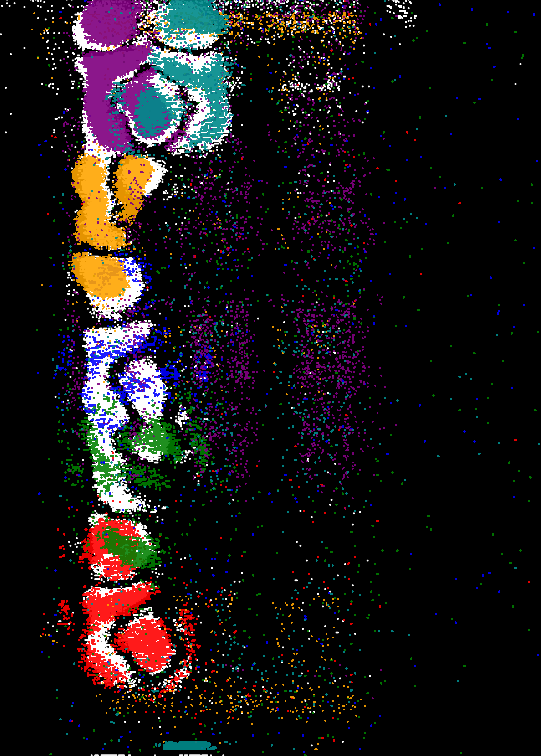}
        \label{subfig:laguna-sll-layout}
    }

    \begin{tikzpicture}[remember picture, overlay]
        \draw[green,ultra thick] (-3.75,7.15) rectangle (-1.35,12.37);
        \draw[green,ultra thick] (0.82,7.15) rectangle (3.22,12.37);
        \draw[green,ultra thick] (-3.75,0.75) rectangle (-1.35,5.97);
        \draw[green,ultra thick] (0.82,0.75) rectangle (3.22,5.97);
    \end{tikzpicture}
    
    \caption{By individually toggling groups of Laguna registers in different configurations, we can use EOFM to make out the layout of individual Laguna sites and drivers. In (a), we highlight the areas that light up when six SLLs in a single Laguna site are toggled. In (b) we redo the same and color-code four different Laguna sites that make up a Laguna tile. These steps are repeated in (c) and (d) for individual SLLs in a single Laguna site.}
    \label{fig:laguna-readout}
\end{figure}

Analogous to how we located specific Laguna tiles within the layout by activating and deactivating blocks of ring oscillators, at higher magnification we further identify specific Laguna registers within a Laguna tile by setting individual SLLs to toggle while holding the rest at a steady value. By this process, we are able to reconstruct the layout of Laguna sites (Figure~\ref{subfig:laguna-site-layout}) and even identify the drivers of individual wires (Figure~\ref{subfig:laguna-sll-layout}).

The entire process of starting from an unknown chip to locating points of interest takes a couple of hours, however, once located the unique structure of SLL drivers makes it possible to visually identify Laguna columns within minutes with no EOFM required. In Figure~\ref{subfig:laguna-columns}, a Laguna column can be seen with Laguna tiles separated vertically by white bars. Note that identifying individual elements in the FPGA fabric in a similar fashion is not possible due to their small feature size.

\subsection{Reading waveforms with EOP}
\label{subsec:results-EOP}
\begin{figure}[htbp]
    \centering
    \begin{tikzpicture}
        \node[anchor=south west,inner sep=0] (image) at (0,0) {\includegraphics[width=0.99\linewidth]{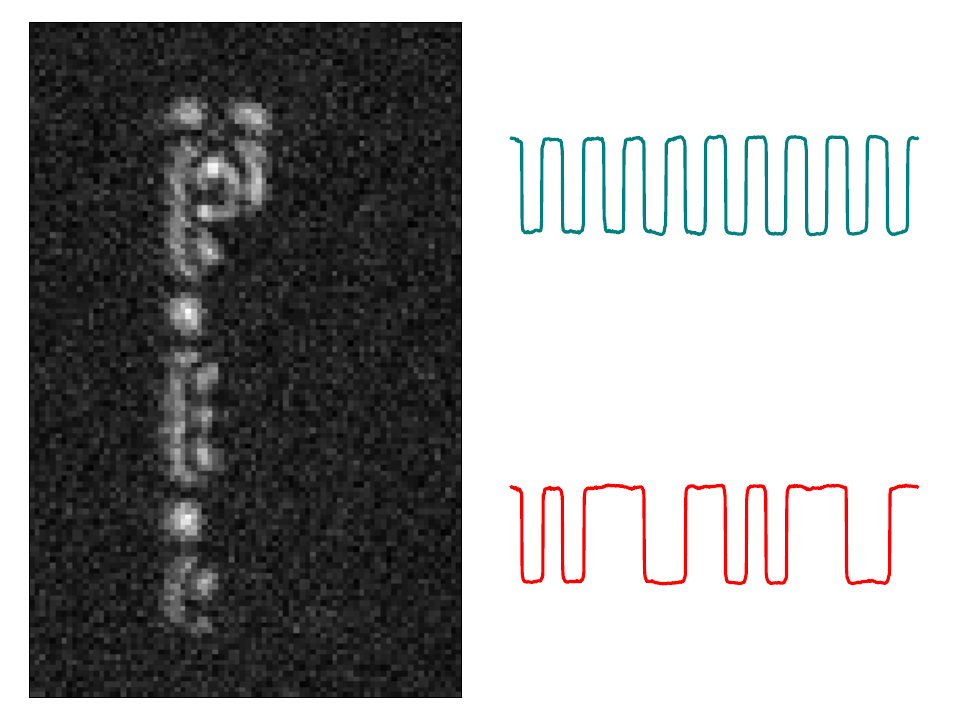}};
        
        \draw[teal,ultra thick] (4.5,4) rectangle (8.6,6);
        \draw[teal,ultra thick,<-] (2.3,5) -- (4.5,5) {};
        \draw[red,ultra thick] (4.5,0.75) rectangle (8.6,2.75);
        \draw[red,ultra thick,<-] (1.9,1.2) -- (4.5,1.75) {};
        
    \end{tikzpicture}
    \caption{Mappings such as that in Figure~\ref{fig:laguna-readout} can be used to locate wires of interest, allowing one to probe specific data lines. The tips of arrows in the figure mark the exact location of probes with the extracted waveforms on the right.}
    \label{fig:eofm-eop}
\end{figure}

\begin{figure*}
    \centering
    \includegraphics[width=0.98\textwidth]{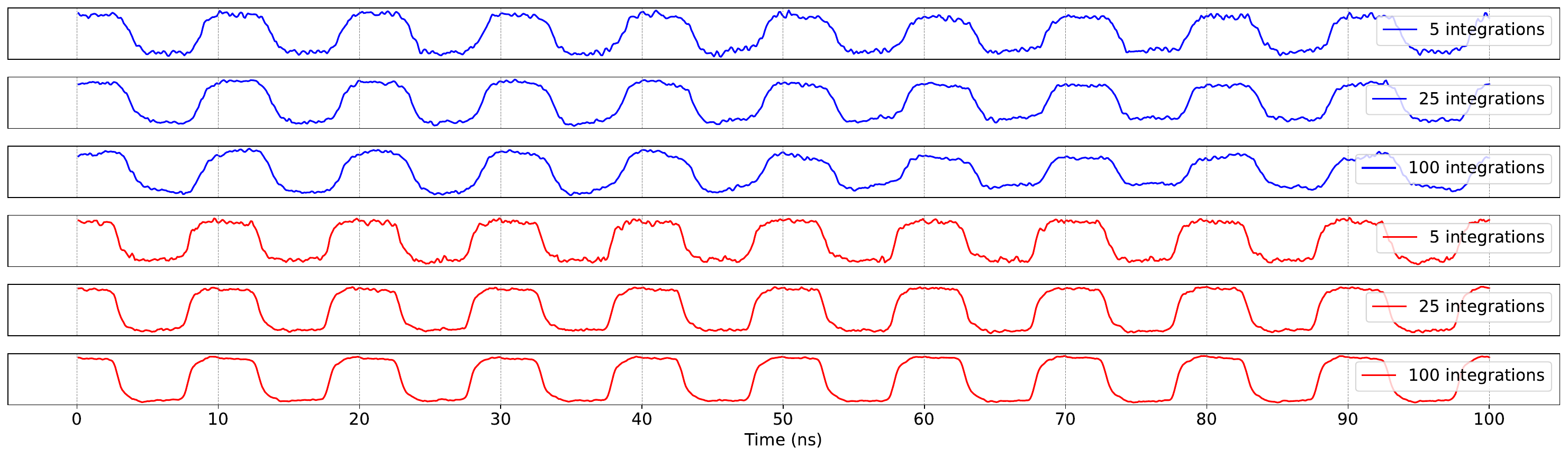}
    \caption{Probed waveforms of fabric (blue) and Laguna (red) registers using 5, 25 and 100 integrations.}
    \label{fig:eop-snr-comp}
\end{figure*}

Once we locate individual elements with EOFM, we use EOP to read out waveforms from the transistors. Figure~\ref{fig:eofm-eop} shows how positioning the laser probe at specific color-coded points in Figure~\ref{subfig:laguna-sll-layout} in a Laguna site can be used to extract data from specific individual data lines. While this is also achievable in fabric, locating the points to probe requires much more effort as longer, higher-resolution scans are needed. Figure~\ref{fig:eop-snr-comp} shows the results of probing fabric (in blue) and Laguna registers (in red) at 5, 25 and 100 integrations. For both traces, the quality of the signal increases with number of integrations, however, the fabric waveform is markedly worse as evident from its noisier flat sections.

\section{Delay-Based Sensing and Mitigation}
\label{sec:delay-sensors}
The previous section demonstrated the ease of locating and probing individual interposer signals. A natural way to protect against probing then might involve monitoring propagation delays of signals one wishes to protect in hopes of being able to detect probing when it occurs. When probed, whether by a physical or a contactless probe, the delay of the probed wire increases due to the increase in capacitance or temperature, respectively, triggering the monitoring sensor. Hence, in this section, we quantify the effect of contactless probing on the delay of a single interposer wire. We examine two types of sensors: a sensor based on two phase-shifted clocks and a Time-to-Digital Converter (TDC).

\subsection{Differential Sensing Circuits}
Probing a specific spot on a die involves pointing a laser at the area of interest. If the wavelength of the light used is shorter than 1.1$\mu$m, photoexcited electrical carriers might be generated in the silicon as electrons from a lower energy band are excited to a higher one, generating electrical current and potentially injecting logical faults into circuits. The light source in the PHEMOS-X however only has the wavelength of 1.3$\mu$m, meaning its photons have less energy than the bandage of silicon and thus cannot induce currents; rather, the point being probed only experiences localized heating that leads to an increase in propagation delay of signals which can be picked up using on-chip sensors.

Apart from probing, the temperature of the chip also varies due to environmental factors. In a steady stream of delay measurements this manifests as a drift in delay over time, which makes it hard to discern the specific impact of probing. For this reason in the following sections we do not consider absolute delays of wires, but rather differential delays. That is, we record delays from two wires placed far apart, such that the probe only affects the first wire while environmental factors affect both. In other words, the second wire acts as a control to normalize the delay against environment. Accordingly, we define differential delay as the difference between the probed wire delay and the delay of the control wire.

\begin{figure}
    \centering
    \subfloat[Dynamic phase-shifting-based sensor]{
        \includegraphics[width=0.47\textwidth]{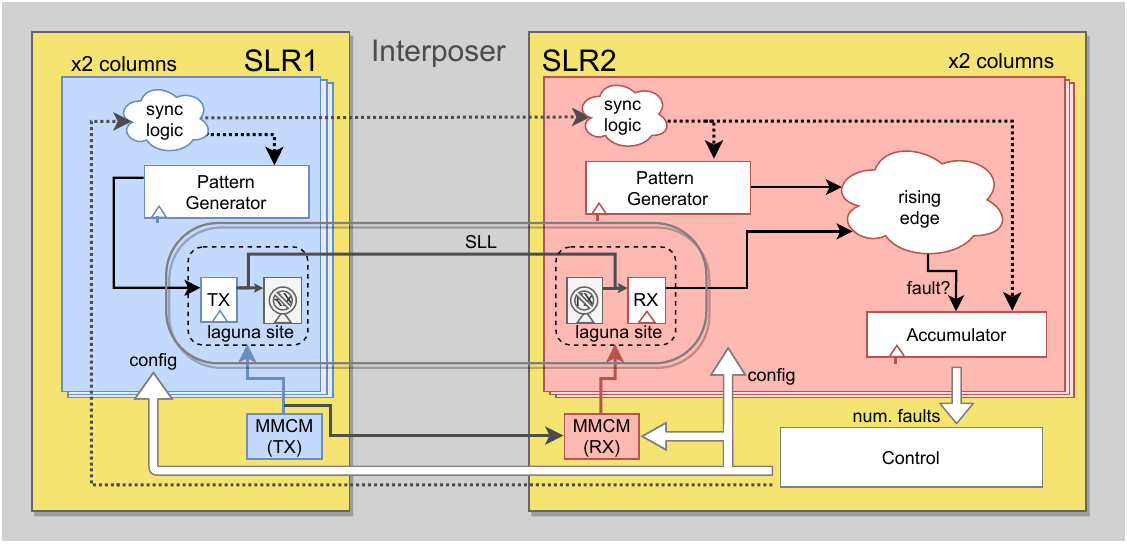}
        \label{fig:phase-shifting-diagram}
    }
    \hfill
    \subfloat[TDC-based sensor]{
        \includegraphics[width=0.47\textwidth]{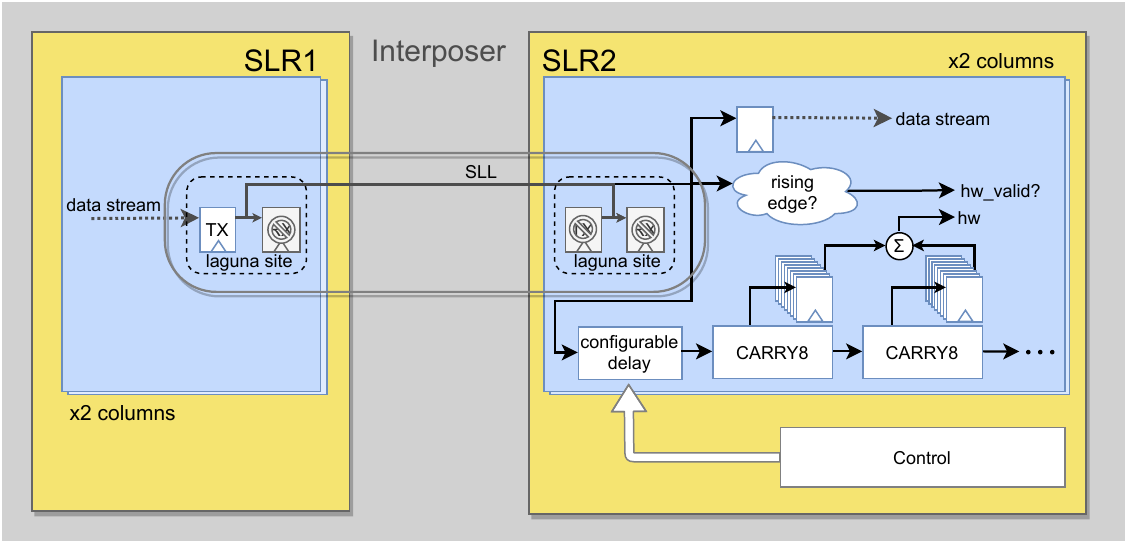}
        \label{fig:shadow-tdc-diagram}
    }
    \caption{{High-level block diagram of the two delay sensors.}}
\end{figure}

We report results from two sensors: a phase-sweeping-based sensor and a TDC. High-level block diagrams of each sensor are shown in Figures~\ref{fig:phase-shifting-diagram} and~\ref{fig:shadow-tdc-diagram}, respectively. In the following sections we describe their working mechanisms and discuss their pros and cons.

\subsubsection{Phase-Sweeping Sensor}
The phase-sweeping sensor is the same one detailed in~\cite{deric2022know}. To recount, in a source synchronous clocking scheme the transmitting chiplet forwards its clock along with the data to the receiving chiplet. The receiving chiplet includes a mechanism like a delay-locked loop (DLL) or a phase-locked loop (PLL) to adjust the phase of the incoming clock and allow for accurate data line sampling. In our scheme, data is transmitted using custom placed Laguna registers on either end of an SLL while the incoming clock is fed into a mixed-mode clock manager (MMCM) before being used for sampling. 

To measure the propagation delay of an SLL we custom place two registers at the boundary of two SLRs. Using the dynamic phase shifting capability of the MMCM we then sweep the phase of the clock driving the receiving register in relation to the clock driving the transmitting register. While this is done, the transmitting register sends a pattern of rising and falling edges that is also reproduced on the receiving side and used to check for correctness of the sampled values. Depending on the phase relationship between the transmitting and receiving clocks, values sampled by the receiving register will either be correct or incorrect. Initially, assuming a large enough phase difference no values will be sampled incorrectly, however, as the phase difference is reduced (in increments of 14.286 ps) and the setup time of the circuit is violated, with an increasing probability the receiving register will sample values incorrectly. The phase difference (in ps) at which the transmitted rising edge is sampled correctly 50\% of the time is taken as the propagation delay of the wire.

Although not the primary focus in our work, one downside of this design is its significant overhead. While the delay of a wire is being measured, the line cannot be used to transmit any useful data as it must transmit the predefined pattern. Similarly, additional clock cycles are wasted on waiting for the phase shift to take effect and logic synchronization across the two clock domains.

\subsubsection{Time-to-Digital Converter}
The timing penalty for going through the interposer and crossing chiplet boundaries is high, which is why AMD/Xilinx advises that large designs that will utilize multiple SLRs be highly pipelined. However, signals crossing chiplet boundaries do not necessarily have to start or end at Laguna registers. AMD/Xilinx imposes three restrictions on placement of RX registers: 1) if there is only one load register, that register can be assigned to a Laguna RX site; 2) if there are multiple loads, no load register can be assigned to a Laguna RX site; 3) all loads must reside in the opposite SLR. A simple way of measuring delay involves setting up a time-to-digital converter (TDC) across the chiplet boundary.

The full design consists of a single launch register that is custom placed to a Laguna TX register. The register drives its associated wire that crosses through the interposer and goes into the other chiplet, however, the signal does not terminate at the corresponding Laguna RX flop, rather it leaves its Laguna site and is sampled by a register in the fabric of the FPGA before continuing into a TDC. The TDC is implemented in typical fashion: the incoming signal is first passed through a series of LUTs whose outputs can be individually selected to feed into the delay line for calibration purposes. The delay line is implemented using 30 CARRY8 primitives and has a total of 240 taps. In this setup, a single clock is used to clock all sequential elements. 

Apart from being simple, the main benefit of this approach over the phase-sweeping design is that no time multiplexing of the wire is needed to measure its delay. The first flop that samples the incoming data is followed by two more registers that delay the signal and are used to generate the valid signal for the Hamming weight output. We deliberately utilize a long carry chain to simplify calibration, however by careful design of the tunable delay component and the delay line the design can be further optimized. Furthermore, we argue that our TDC can be substituted with any other FPGA delay sensor if desired.

\subsection{Delay Change}

\begin{figure}
    \centering
    \subfloat[Change in differential delay.]{
        \centering
        \includegraphics[width=0.95\linewidth]{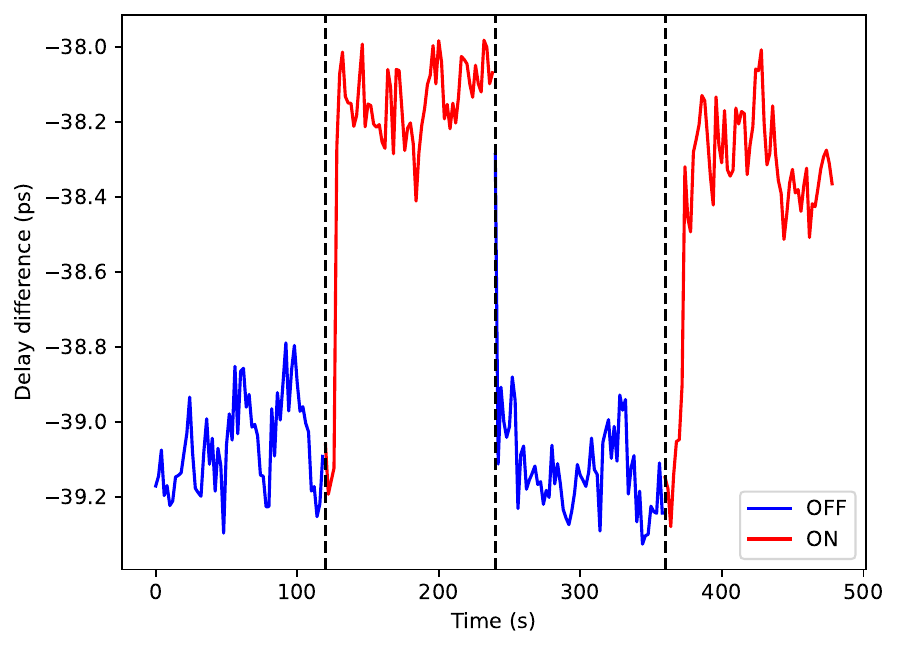}
        \label{subfig:phase-timeseries-diff}
    }
    \hfill
    \subfloat[Differential delay distributions when the laser is on/off.]{
        \centering
        \includegraphics[width=0.95\linewidth]{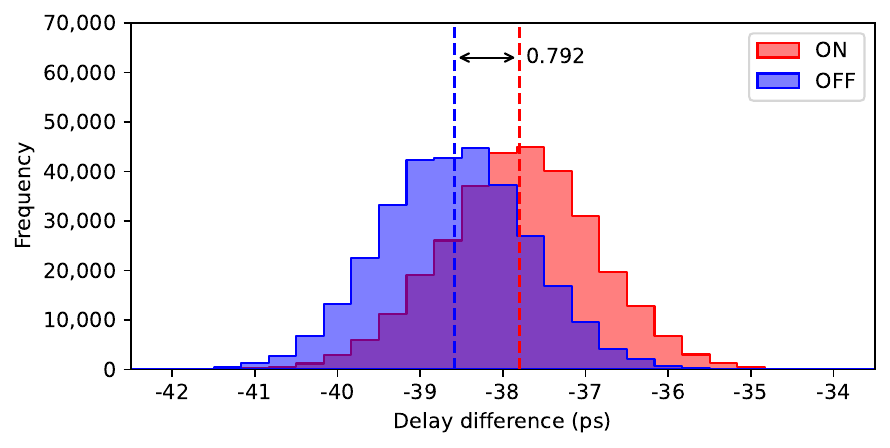}
        \label{subfig:phase-probed-hist}
    }
    
    \caption{Effects of probing a Laguna driver on interposer wire delay measured using the phase-shifting sensor. (a) shows the moving average of delay with a two-second window.}
    \label{fig:phase-delay-measurements}
\end{figure}

Figure~\ref{fig:phase-delay-measurements} shows the results of probing an SLL with the phase-sweeping sensor while the laser is set to 100\% power. The laser is toggled every two minutes, starting with it off. When the laser is off, the delay difference is -39.090~ps (the difference is negative as the control wire is slower than the probed wire). Once the laser is turned on, the delay of the probed wire increases, resulting in an average differential delay of -38.298~ps, a 0.792~ps jump. Figure~\ref{fig:delay-change-vs-laser-power} additionally shows the delay change at 75\%, 50\% and 25\% power using the same sensor.

\begin{figure}
    \centering
    \subfloat[Change in differential Hamming weight.]{
        \centering
        \includegraphics[width=0.95\linewidth]{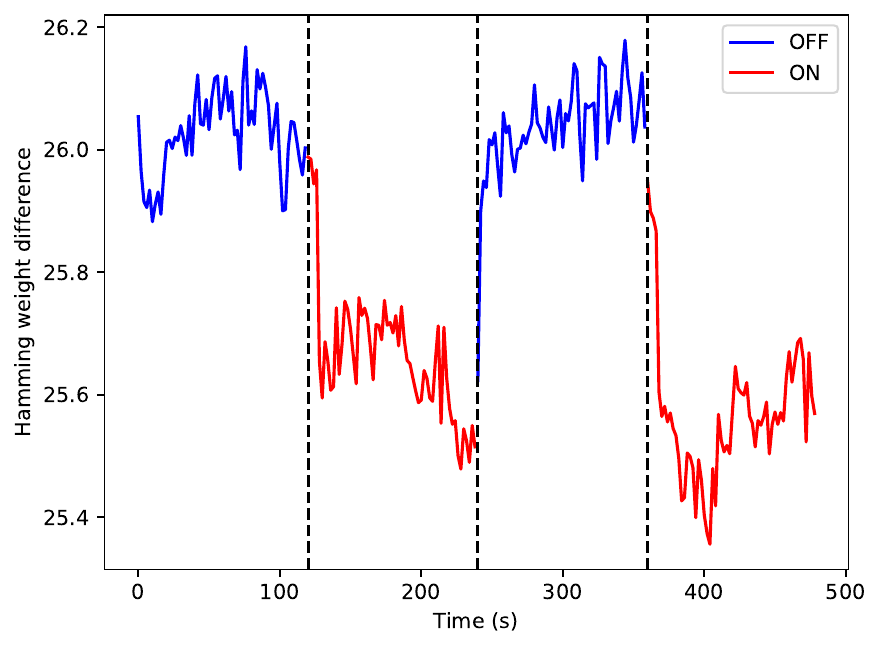}
        \label{subfig:tdc-timeseries-diff}
    }
    \hfill
    \subfloat[Differential Hamming weight distributions when the laser is on/off.]{
        \centering
        \includegraphics[width=0.95\linewidth]{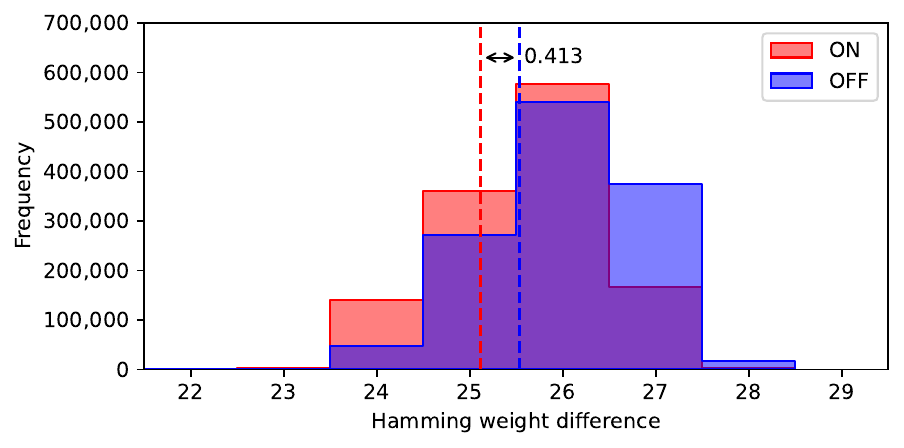}
        \label{subfig:tdc-probed-hist}
    }
    
    \caption{Effects of probing a Laguna driver on differential Hamming weight measurements made using TDCs. (a) shows the moving average with a two-second window.}
    \label{fig:tdc-hw-measurements}
\end{figure}

Figure~\ref{fig:tdc-hw-measurements} summarizes the results from the TDC sensor where Hamming weight is used in place of delay measurements. Initially, the difference between the output of the two sensors is positive, but when the SLL feeding into the first sensor is probed, the change in propagation delay causes its output to drop, resulting in a dip in differential measurements. At 100\% laser power, this dip is observed to be 0.413 taps. Assuming that the true delay is 0.792~ps as measured by the phase-sweeping sensor, each tap can then be calculated to add 1.917~ps of delay, which is reasonably consistent with 2.75~ps that AMD/Xilinx provides as its conservative estimate for timing in Vivado.

\subsection{Obstacles to Reliable Detection}
\begin{figure}
    \centering
    \includegraphics[width=0.45\textwidth]{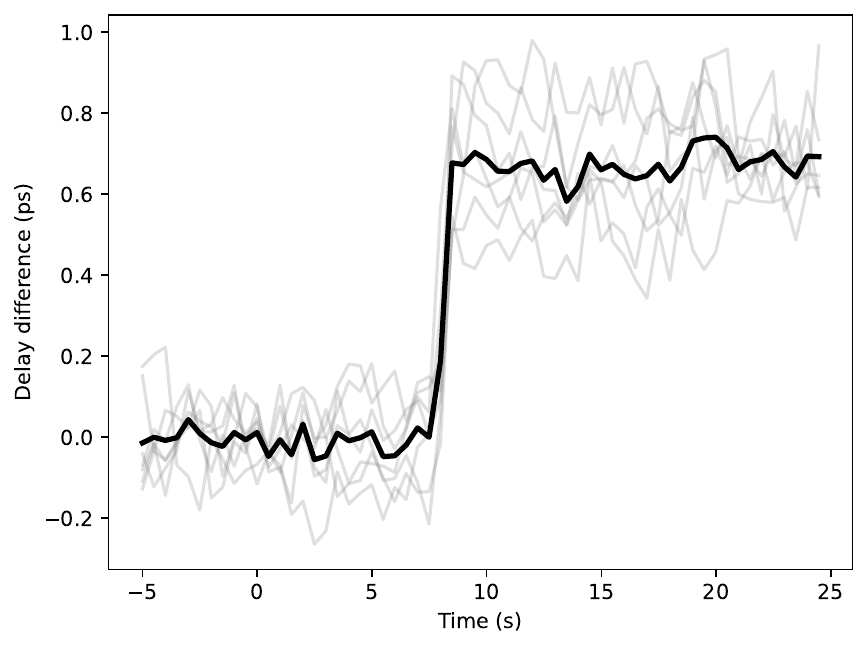}
    \caption{{Plot illustrating the change in differential delay across multiple measurements after turning on the laser probe (set to 100\% power). The solid black line represents the mean of all measurements.}}
    \label{fig:sensor-change-from-probe}
\end{figure}

\begin{figure}
    \centering
    \includegraphics[width=0.45\textwidth]{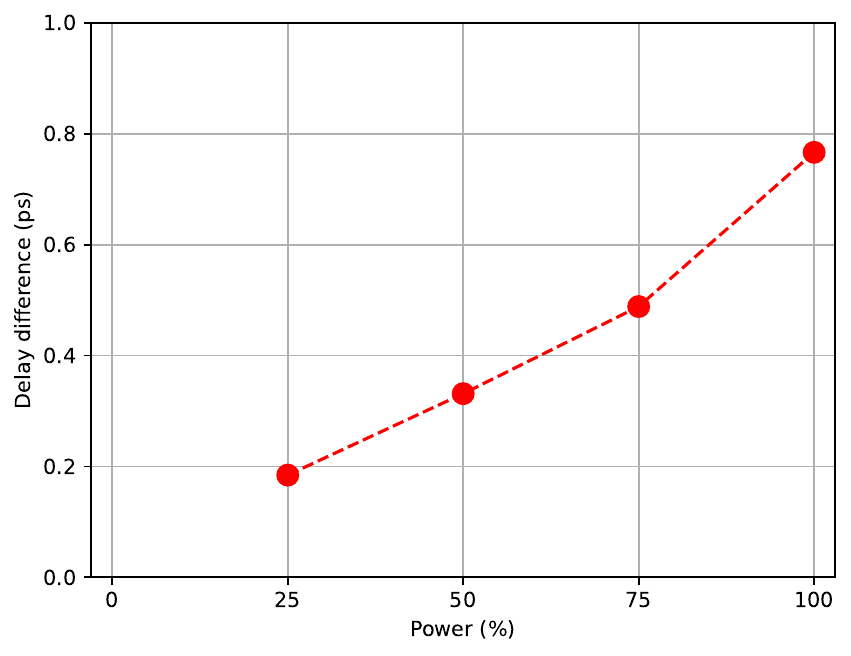}
    \caption{{Delay change of an SLL at different laser power settings. Expectedly there is a roughly linear relationship between the two.}}
    \label{fig:delay-change-vs-laser-power}
\end{figure}

Our measurements in the previous subsection show that the effects of probing can be observed by delay sensors, though conclusively detecting that a line is being probed remains an open challenge as the delay at most increases by 0.8ps and does so steadily over a period of one second as shown in Figure~\ref{fig:sensor-change-from-probe}. The challenge is then to distinguish whether this increase is due probing or other potentially factors such as temperature changes due to local logic switching. Furthermore, Figure~\ref{fig:delay-change-vs-laser-power} shows a linear relationship between laser power and delay change, meaning for easily probed signals such as SLL drivers the adversary could further reduce the laser power, making any delay changes due to probing even harder to distinguish from noise.

\subsection{Efficient Deployment of In-Situ Sensing}
Though with laser probing one can probe any point in the chip, as discussed in Section~\ref{sec:laser-probing} it is easier to locate and probe chiplet drivers than fabric registers. Likewise, a limitation of contactless probing is that the party doing the probing needs the ability to replay the data stream they wish to extract as EOP depends on integrating many traces to reconstruct the probed waveform. In real attack scenarios, this is typically achieved via resets.

One way of partially mitigating the issue involves masking the data using a small key that is sent serially over wires instrumented with sensing capabilities. A simple and cheap mechanism to achieve this is for the transmitting chiplet to include a true random number generator (TRNG) that generates a one-time-pad to XOR with the outgoing data stream. Using the pad the receiving chiplet is able to decode the data by combining incoming data lines with the pad stream. Note that reusing the pad for all data lines is generally considered a poor practice, however the goal of the system is not secure encryption but rather masking the data. As the TRNG never regenerates the same stream of values, even if the system is forced into replaying data the actual values on the interconnect wires will be random, preventing any adversary from being able to integrate multiple waveforms and extract the actual data.

\section{Conclusion}
\label{sec:conclusion}
As SoC designs become increasingly common due to their performance and power benefits, chiplets are expected to continue to play an important role as they offer a way to get around reticle limits. Apart from packaging and test challenges, chiplets also bring unique security challenges. Our work in this paper shows that their large drivers that power inter-chiplet wires make them an easy and attractive target for probing. We also implement two types of cross-die delay sensors and evaluate their ability to detect contactless probing. Our findings indicate that while the sensors are capable of measuring the small delay changes due to contactless probing, reliably distinguishing probing from environmental fluctuation remains a challenge and more work should be conducted to address the vulnerability highlighted by this paper. In the meantime, techniques like simple masking can help protect against probing attacks.

\section*{Acknowledgment}
This research was supported in part by National Science Foundation Grants CNS-1902532 and CNS-2150123, and by a Research and Development (R\&D) grant from the Massachusetts Technology Collaborative.

\bibliographystyle{ieee}
\bibliography{main}

\end{document}